\documentclass[11pt,twoside]{article}
\usepackage{asp2004}\usepackage{psfig}\usepackage{epsf}
\usepackage{graphics}\usepackage{lscape}
\makeindex 
\begin{document}
\title{Kepler Equation for the Compact Binaries under the 
Spin-spin Interaction}
\author{Z. Keresztes \& B. Mik\'{o}czi}
\affil{Departments of Theoretical and Experimental Physics, 
University of
Szeged, Szeged 6720, Hungary}

\begin{abstract} In this work we consider compact binaries on 
eccentric orbit under the spin-spin
interaction. Using the post-Newtonian formalism, the binaries 
undergo a perturbed Keplerian 
motion. Here we investigate only the radial motion and derive the 
contribution of the second 
post-Newtonian order spin-spin effect to the solution.
\end{abstract}

\section{Introduction} 

The solution to the relativistically perturbed two body problem 
\citep{Damour} has 
received important applications since the discoveries of compact 
binary pulsars \citep{Hulse}. The predictions of the general 
theory of relativity for 
the motion of such systems were 
verified to a high degree. Nowadays compact binary systems 
composed of neutron stars and/or black 
holes are considered among the most promising sources for the 
Earth-based gravitational wave 
observatories. Before the final coalescence there is a regime 
where a post-Newtonian (PN) 
description of the motion and of the gravitational radiation is 
suitable.

   The dynamics of a two-body system without spins to 2PN orders 
\citep{Damour2} and 
then the perturbation induced by the spin-orbit interaction 
\citep{Schafer, Wex} was 
solved using suitable quasi-Keplerian parametrizations. A 
systematic treatment \citep{Gergely3} has yielded such 
parametrizations 
(the true and eccentric anomaly 
parametrizations) for the radial motion of a wide class of 
perturbed Keplerian motions, including 
the generic perturbing force of Brumberg \citep{Brumberg}. These 
parametrizations have convenient 
features over those introduced by \citep{Ryan, Rieth}.

   The first imprint of the presence of the spins in a compact 
binary are the spin-orbit effects 
\citep{Rieth, Gergely2}. At 
the 2PN order, also spin-spin 
effects appear \citep{Gergely1}. Therefore it would be desirable 
to 
have the corresponding 
parametrization of the orbit available. The main impediment in 
the parametrization of the 
perturbed Keplerian motion with the inclusion of the spin-spin 
effects is that the magnitude of 
the orbital angular momentum is not conserved, even in the 
absence of gravitational radiation. 
This feature of the spin-spin type perturbation is outside the 
framework settled in \citep{Gergely3}. The radial motion in 
terms of true 
anomaly parametrization was 
solved \citep{Gergely1}. However the complete dynamics with the 
inclusion of angular degrees of 
freedom was not described yet.

\section{The Radial Motion}

The motion of the binary under the influence of the spin-spin 
interaction is governed by the 
Lagrangian \citep{Kidder}:
\[
\mathcal{L=L}_{N}+\mathcal{L}_{SS},
\]
where
\begin{eqnarray*}
\mathcal{L}_{N} &=&\frac{\mu \mathbf{v}^{2}}{2}+\frac{Gm\mu }{r},
\\
\mathcal{L}_{SS} &=&\frac{G}{c^{2}r^{3}}\left[ \left( 
\mathbf{S}_{\mathbf{1}}
\mathbf{S}_{\mathbf{2}}\right) -\frac{3}{r^{2}}\left( 
\mathbf{rS}_{\mathbf{1}
}\right) \left( \mathbf{rS}_{\mathbf{2}}\right) \right].
\end{eqnarray*}
The perturbed motion is characterized by the conservation of the 
total energy
\[
E=E_{N}+E_{SS},\quad E_{N}=\frac{\mu \mathbf{v}^{2}}{2}-
\frac{Gm\mu }{r},\quad E_{SS}=-\mathcal{L}_{SS}.
\]
and of the total angular momentum vector
\[
\mathbf{J}=\mathbf{L}+\mathbf{S},
\]
where $\mathbf{L=L}_{N}=\mu \left( \mathbf{r}\times 
\mathbf{v}\right) $ is the orbital angular 
momentum and $\mathbf{
S=S}_{\mathbf{1}}\mathbf{+S}_{\mathbf{2}}$ is the total spin.
Using the equation of the motion the magnitude of the orbital 
angular momentum is not constant: it 
changes according to (2.14) of \citep{Gergely1}. However the 
secular evolution of the orbital 
angular momentum vector \citep{Barker} is a 
precessional motion. Consequently, there 
is no change in the magnitude of the orbital angular momentum 
over one radial period due to the 
spin-spin interaction. Using the procedure given in 
\citep{Gergely1} the magnitude of the orbital 
angular momentum can be written in the following form:
\begin{eqnarray*}
L=\overline{L}+\delta L\left( \chi \right),
\end{eqnarray*}
where the first term is the angular average of the magnitude of 
the orbital angular momentum over 
one radial period and the second appears due to the spin-spin 
interaction. $\chi $ is the true 
anomaly parameter in the Newtonian limit. We can characterize the 
radial motion by the total 
energy and $\overline{L}$ instead of $L$. 

The expression of the energy and orbital angular momentum in 
terms of spherical coordinates give
the {\it radial equation}, which for the accuracy needed here is:
\[
\dot{r}^{2}=\frac{2E}{\mu 
}+\frac{2Gm}{r}-\frac{\overline{L}^{2}}{\mu
^{2}r^{2}}-\frac{2\overline{L}\delta L}{\mu ^{2}r^{2}}+\frac{
GS_{1}S_{2}\alpha }{c^{2}\mu r^{3}},
\]
where $\delta L$ and $\alpha $ is given in \citep{Gergely1}. To 
solve the radial equation we 
introduce the generalized true and eccentric anomaly 
parametrization \citep{Gergely1}. The relation 
between the two parametrizations:
\begin{eqnarray*}
\xi =2\arctan \left( \sqrt{\frac{1-e_{r}}{1+e_{r}}}\tan 
\frac{\chi }{2}
\right).
\end{eqnarray*}
Here the eccentric anomaly parameter is denoted by $\xi $ and the 
true anomaly parameter by $\chi 
$. Using these parametrizations the integration of the radial 
equation emerges after a long 
computation. The solution of the radial equation is 
\begin{eqnarray*}
r &=&a\left( 1-e_{r}\cos\xi \right), \\
n\left( t-t_{0}\right)  &=&\xi -e_{t}\sin\xi 
+\frac{f_{t}}{c^{2}}\sin \left[ \chi +2\left( \psi 
_{0}-\overline{
\psi }\right) \right],
\end{eqnarray*}
where
\begin{eqnarray*}
a_{r} &=&\frac{Gm\mu }{-2E}\left[ 
1+\frac{ES_{1}S_{2}}{c^{2}m\overline{L}^{2}
}\left( \alpha _{SS}+\beta _{SS}\right) \right], \\
e_{r} &=&\frac{\overline{A}}{Gm\mu }\left\{ 
1-\frac{ES_{1}S_{2}}{c^{2}m
\overline{L}^{2}\overline{A}^{2}}\left[ \left( G^{2}m^{2}\mu 
^{2}+\overline{A
}^{2}\right) \alpha _{SS}+\overline{A}^{2}\beta _{SS}\right] 
\right\}, \\
n &=&\frac{2\pi }{T}=\frac{1}{Gm}\left( \frac{-2E}{\mu }\right) 
^{3/2}, \\
e_{t} &=&\frac{\overline{A}}{Gm\mu }\left[ 
1-\frac{ES_{1}S_{2}G^{2}m\mu
^{2}\alpha _{SS}}{c^{2}\overline{L}^{2}\overline{A}^{2}}\right],
\\
f_{t} &=&-\left( \frac{-2E}{\mu }\right) ^{3/2}\frac{\mu 
S_{1}S_{2}}{c^{2}m
\overline{A}\overline{L}}\sin \kappa _{1}\sin \kappa _{2},
\end{eqnarray*}
with
\begin{eqnarray*}
\alpha _{SS} &=&3\cos\kappa _{1} \cos\kappa _{2}-\cos\gamma, \\
\beta _{SS} &=&\sin\kappa _{1} \sin\kappa _{2} \cos 2\left( \psi 
_{0}-\overline{\psi }\right).
\end{eqnarray*}
The magnitude of the Laplace-Runge-Lenz vector $\overline{A}$ 
belongs to a Keplerian motion 
characterized by the energy $E$ and $\overline{L}$. The $\kappa 
_{i}$ and $\gamma $ are relative 
angles of the orbital angular momentum vector and the spins of 
the bodies. The angles $\psi _{0}$ 
and $\psi _{i}$ are subtended by the intersection line of the 
planes perpendicular to 
$\mathbf{L}_{N}$ and $\mathbf{J}$ with the position of periastron 
line and the projections of the 
spins in the plane of the orbit. All angles are shown in Fig. 1 
in \citep{Gergely2}.
 
\section{Concluding Remarks}

We have presented the solution of the radial motion of a 
Newtonian problem perturbed by spin-spin 
interaction. In part, the orbital elements are characterized by 
the energy $E$ and the angular 
average of the magnitude of the orbital angular momentum 
$\overline{L}$. The magnitude of the 
spins, angles between the spins and angles between the spins and 
the orbital angular momentum also 
appear in the characterization of the motion. During one radial 
period the angles $\kappa _{i}$, 
$\gamma $, $\psi _{i}$ and $\psi _{0}$ can be considered 
constants \citep{Gergely2}, thus the orbital elements are 
constants as 
well.

This work was supported by OTKA grants no. T046939 and TS044665.
  
{}

\end{document}